# DUST-BOUNDED ULIRGS? MODEL PREDICTIONS FOR INFRARED SPECTROSCOPIC SURVEYS

N. P. Abel[1], C. Dudley[2], Jacqueline Fischer[3], S. Satyapal[2], & P. A. M. van Hoof[4]

[1]College of Applied Science, University of Cincinnati, Cincinnati, OH, 45206
npabel2@gmail.com

[2]Department of Physics and Astronomy, George Mason University, Fairfax, VA 22030-4444, satyapal@physics.gmu.edu, dudley@vivaldi.nrl.navy.mil

[3]Naval Research Laboratory, Remote Sensing Division, Code 7211, Washington, DC 20375, Jackie.Fischer@nrl.navy.mil

[4]Royal Observatory of Belgium, Ringlaan 3, 1180 Brussels, Belgium;
p.vanhoof@oma.be

## Abstract

The observed faintness of infrared fine-structure line emission along with the warm far-infrared (FIR) colors of ultraluminous infrared galaxies (ULIRGs) is a long-standing problem. In this work, we calculate the line and continuum properties of a cloud exposed to an Active Galactic Nucleus (AGN) and starburst spectral energy distribution (SED). We use an integrated modeling approach, predicting the spectrum of ionized, atomic, and molecular environments in pressure equilibrium. We find that the effects of high ratios of impinging ionizing radiation density to particle density (i.e. high ionization parameters, or $U$) can reproduce many ULIRG observational characteristics. Physically, as $U$ increases, the fraction of UV photons absorbed by dust increases, corresponding to fewer photons available to photoionize and heat the gas, producing what is known as a "dust-bounded" nebula. We show that high $U$ effects can explain the "[C II] deficit", the ~1 dex drop in the [C II] 158μm/FIR ratio seen in ULIRGs when compared to starburst or normal galaxies. Additionally, by increasing $U$ through increasing the ionizing photon flux, warmer dust and thus higher IRAS $F$(60μm)/$F$(100μm) ratios result. High $U$ effects also predict an increase in [O I]63μm /[C II] 158μm and a gradual decline in [O III] 88μm /FIR, similar to the magnitude of the trends observed, and yield a reasonable fit to [Ne V]14μm /FIR ratio AGN observations.

## 1 Introduction & Background

The study of infrared galaxies discovered with IRAS appears to show a very intimate relation between powerful infrared emission and the evolution of galaxies in merger events. However, the same dust which is responsible for the emission is an obstacle to traditional optical spectroscopic diagnostics. The role played by interactions and mergers of gas rich galaxies and the associated infrared-luminous galaxies in galaxy evolution is of interest now and in the near future as infrared and sub-mm surveys obtained by missions such as the *Spitzer Space Telescope*, the *Submillimeter Observatory For Infrared Astronomy* (*SOFIA*), and the *Herschel Space Observatory* become available. One of the most important and intriguing discoveries of the *Infrared Space Observatory (ISO)* mission was the discovery that in ultraluminous infrared galaxies (ULIRGs), galaxies with infrared luminosities $L_{IR} \geq 10^{12} L_\odot$, the far-infrared (FIR) fine-structure line luminosities from both atomic and ionized gas are as much as an order of magnitude fainter relative to the FIR luminosity, than in less luminous infrared bright galaxies powered by either AGN or starbursts (Luhman et al. 1998; Luhman et al. 2003, henceforth L03; Fischer et al. 1999). This effect is best characterized for the [C II] 158µm line, typically one of the strongest emission lines in galaxies, for which it is known as the "[C II] Line Deficit"(L03). In an *ISO Long Wavelength Spectrometer (LWS)* survey of a sample of 15 nearby ULIRGs, L03 found an order of magnitude deficit in the strength of the [C II] 158µm line relative to the far-infrared dust continuum emission[1].

Other observational trends were established with the *ISO LWS*, including the trend of decreasing [C II] 158µm /FIR and increasing [O I]63 µm /[C II] 158 µm ratio with increasing IRAS $F_{60\mu m}/F_{100\mu m}$ [henceforth $F(60)/F(100)$] color ratio (Malhotra et al. 1997, 2001, L03), consistent with the [C II] deficit in ULIRGs, since ULIRGs have relatively high IRAS F(60)/F(100) ratios compared with normal galaxies.

What is responsible for the observed line deficits and color differences between ULIRGs and other galaxies? Malhotra et al. (1997, 2001) attributed many of the effects seen in warm (high $F(60)/F(100)$ ratio) galaxies including ULIRGs, to high ratios of $G_0/n_H$ in the neutral galactic medium, where $G_0$ is the incident far-UV (FUV) 6 – 13.6 eV radiation field, normalized to the average local interstellar radiation field (Habing 1968), and $n_H$ is the total hydrogen gas density. A more general idea proposed in L03 to explain deficits from lines that trace the ionized medium as well as the neutral medium, is that ULIRGs are characterized by a higher ionization parameter (*U*) than in normal and starburst galaxies, where the ionization parameter *U* is defined as the dimensionless ratio of the incident ionizing photon density to the hydrogen density:

$$U \equiv \frac{\varphi_H/c}{n_H} = \frac{Q(H)}{4\pi R^2 n_H c} \tag{1}$$

---

[1] FIR(42.5–122.5µm), as measured by the IRAS Faint Source Catalog fluxes and the prescription given in the appendix of Helou et al. (1988)

where $\phi_H$ is the flux of hydrogen ionizing photons striking the illuminated face of the cloud per second, $c$ is the speed of light, $Q(H)$ is the number of hydrogen ionizing photons striking the illuminated face per second, and $R$ is the distance of the ionizing source to the illuminated face. In a grain free Stromgren H$^+$ region slab of length $L$, $N(H^+) = n_H L = U c / \alpha_B(T) \approx U \times 10^{23}$ where $\alpha_B(T)$ is the case B hydrogen recombination coefficient (Storey & Hummer 1995). With dust present, theoretical calculations predict, for $U > 10^{-2}$, a significant fraction of the hydrogen ionizing photons in the H$^+$ region will be absorbed by dust, i.e. the region will be "dust-bounded" (Voit 1992, Bottorff et al. 1998, Abel et al. 2003). The physical explanation for increased dust absorption of UV photons is that, as $U$ increases, the ratio of ionized to atomic hydrogen increases, since the level of ionization increases with increasing $U$. The increased level of ionization means there are fewer H$^0$ atoms available to absorb Lyman continuum photons, and therefore decreased gas opacity in the H$^+$ region. However, increasing $U$ has no effect on the dust opacity, therefore increasing $U$ leads to an increase in the dust to gas opacity ratio. Increased dust absorption relative to gas photoabsorption decreases the fraction of Lyman continuum photons available for the gas, leading to a reduction in the emission-line fluxes relative to the continuum. However, for a given luminosity, a region which is thick enough will eventually convert all of its energy to dust emission, which means that the total infrared luminosity is the same, whether the absorption occurs in the H$^+$ region or in the atomic region. These effects combine to lower the line-to-bolometric luminosity ratio in high-$U$ environments.

In this paper we explore the hypothesis suggested by L03: that many of the infrared spectral properties of ULIRGs are due to high $U$ effects. Specifically, we seek to determine whether photoionization models that extend beyond the ionized regions into the atomic and molecular media, assuming a set of simple plausible astrophysical assumptions including AGN and starburst spectral energy distributions (SEDs) and pressure balance, can reproduce the observational trends seen in data derived from *IRAS*, *ISO* and *Spitzer*. These models can also be compared with data that will be obtained from surveys expected to be carried out by *Herschel* and *SOFIA* in the coming years. The computational details are presented in Section 2, the detailed results are presented in Section 3, and a comparison of the models with observational data obtained from the literature is presented in Section 4. In Section 5 conclusions drawn from the models and the extent to which they are in agreement with trends found in the data are presented.

## 2 Computational Details

The calculations presented in this work use version C07.02 of the spectral synthesis code Cloudy, last described by Ferland et al. (1998). They include a treatment of the major ionization processes that affect the ionization structure.

These processes include direct photoionization, charge transfer, the Auger effect, autoionization/dielectronic recombination, collisional, Compton recoil, and cosmic ray ionization. In the ionized gas, the most important of these processes is photoionization, although charge transfer ionization/recombination can be important for some species, in particular oxygen (Kingdon & Ferland 1999). Abel et al. (2005) and Shaw et al. (2005) describe recent advances in its treatment of molecular processes, while van Hoof et al. (2004) describe the grain physics. The model parameters are summarized in Table 1. The assumed geometry is a 1D spherical geometry, where the continuum source is completely covered by the gas.

## 2.1 Radiation Field and Stopping Criterion

We consider both AGN and starburst-type SEDs. For the AGN, we used the SED given in Korista et al. (1997) in their study of BLR line intensities. This continuum is characterized by four parameters; the temperature ($T$) of the "Big Bump", the ratio of X-ray to UV flux ($\alpha_{ox}$), the low-energy slope of the Big Bump continuum ($\alpha_{uv}$), and the slope of the X-ray continuum ($\alpha_x$). For our calculations, we use $T = 10^6$ K, $\alpha_{ox} = -1.4$, $\alpha_{uv} = -0.5$, and $\alpha_x = -1.0$. For the starburst input SED, we used a 4 Myr continuous star-formation starburst model taken from the Starburst99 website (Leitherer et al. 1999) as our input SED with a Saltpeter IMF of power law index -2.35 and a star-formation rate of $1\,M_\odot\,\text{yr}^{-1}$. All other parameters were left at their default settings from the Starburst99 website. We define $U$ and $n_H$ at the illuminated face of the cloud. We allow $U$ to vary from $10^{-4}$ to $10^{-0.4}$, in increments of 0.4 dex. This choice of $U$ extends from regions of low ionization to regions where dust absorbs a significant fraction of the Lyman continuum.

Beyond the H$^+$ ionization front, the AGN SED produces an atomic/molecular environment where X-rays control the chemical and thermal structure (henceforth X-ray Dominated Region or XDR; Maloney, Tielens, & Hollenbach 1996). Our choice of AGN SED leads to X-rays contributing to 8% of the total luminosity at the illuminated face. The starburst SED produces a PhotoDissociation Region, or PDR (Tielens & Hollenbach 1985), where the physical processes of the atomic/molecular regions are governed by the Far-UV (FUV) radiation field between 6 – 13.6eV. The intensity of the FUV is often reported in PDR models relative to the interstellar radiation field (ISRF) value of $1.6 \times 10^{-3}$ erg cm$^{-2}$ s$^{-1}$ (Habing 1968), and is denoted as $G_0$. Figure 1 shows the values of $G_0$ and the X-ray intensity (in erg cm$^{-2}$ s$^{-1}$) at the illuminated face as a function of $U$ for our models.

Cosmic rays are also included and are a significant heating source, driving the ion-molecule chemistry deep in the molecular cloud. Primary and secondary cosmic ray ionization processes are treated as described in Appendix C of Abel et al. (2005), with an assumed cosmic ray ionization rate of $5 \times 10^{-17}$s$^{-1}$.

The stopping criterion (corresponding to a particular $A_V$, total hydrogen column density [$N(H_{tot})$], or outer radius) is important to our model results. The thicker we make the cloud, the cooler the dust, which will lower the $F(60)/F(100)$ ratio. Additionally, the thicker the cloud, the larger the optical depths, which can potentially affect the IR fine-structure emission lines such as [O I] 63, 146 µm line (Liseau, Justtanont, & Tielens 2006). At the same time, we have to make our models thick enough to assure that all calculations start off as $H^+$ regions and, by the time the model is stopped, the models are well into PDR/XDR gas. Based on our desire to study variations of IR emission-line and line-to-continuum ratios with $U$ through variations in $\Phi_H$, we decided to stop all calculations at $A_V$ = 100 mag, which corresponds $N(H_{tot})$ = 2×10$^{23}$ cm$^{-2}$. The approximate physical extent (starting radius, stopping radius) of the model, if desired, can be deduced from the definition of $U$ (equation 1), and the parameters listed in Table 1, while the outer radius can be deduced from the hydrogen density in the PDR (Figure 1) and the total column density of the model, given in Table 1. In the case of the outer radius this is only an approximation, as $n_H$ is not constant in the PDR but varies some with depth. For a typical ULIRG, $Q(H)$, ~2×10$^{54}$ s$^{-1}$ (Genzel et al. 1998), the range of $U$ along with the initial hydrogen density corresponds to an inner radius of 30 – 2000 pc. We emphasize that *the model we have adopted is designed to demonstrate the effect of variations in U on the emergent emission line and continuum spectrum to compare to the observational trends*. The actual parameters are not intended to fit data on any individual galaxy.

## 2.2 Chemical Composition and Grain Properties

We assume gas phase abundances consistent with the galactic ISM, which we define as the average abundances taken from the warm and cold phases of Cowie & Songaila (1986) and ξ Oph (Savage & Sembach 1996, Table 5). The oxygen abundance is from Meyers et al. (1998). For the most important species, the abundances by number are He/H = 0.098, C/H = 1.4×10$^{-4}$; O/H = 3.2×10$^{-4}$, N/H = 8×10$^{-5}$, Ne/H = 1.2×10$^{-4}$, Ar/H = 2.8×10$^{-6}$, and S/H = 3.2×10$^{-5}$.

In actuality, gas-phase abundances will vary with depth. As the gas becomes colder, depletion of elements out of the gas phase onto grain surfaces reduce the gas-phase abundance. Additionally, it is possible for grains to be destroyed in the ionized gas, leading to enhanced gas-phase abundances in the $H^+$ region. For simplicity, we neglect any gas-phase abundance enhancements due to dust destruction. This assumption could potentially affect our prediction of fine-structure line intensities in the $H^+$ region, where the gas-phase abundance may be higher than in the cooler PDR/molecular cloud. For this work, the $H^+$ region fine-structure lines involve C, N, Ne, and O, which are not heavily depleted relative to $H^+$ region abundances. For these elements, our assumption of constant gas-phase abundance could potentially underpredict the line emission by a factor of two. We do treat the condensation of $H_2O$, CO, and OH (see

Section 3.1) onto grain surfaces, which is only important when the gas and dust temperature falls below 20 K. These reactions can significantly reduce the gas-phase abundance of C and O deep in our calculations.

The assumed grain size distribution is representative of star-forming regions. The ratio of total to selective extinction, $R_V = A_V/(A_B-A_V)$, is a good indicator of the size distribution of grains (Cardelli et al. 1989). Calzetti et al. (2000) derived an average value for $R_V \sim 4.3$. We therefore use the $R_V = 4$, $b_C = 0$ grain size distribution given in Weingartner & Draine (2001a). The grain abundance is scaled such that $A_V/N(H_{tot}) = 5\times10^{-22}$ mag cm$^2$. As the dust in the H$^+$ region is important in our models, it is worth noting that the ratio of dust absorption at 912Å to $A_V$, $A_{Ly}/A_V$, for our models is 1.75. Additionally, while we include the effects of molecules freezing out onto grains, we do not include the effects of ices on the dust opacity or IR emission.

We also include size-resolved PAHs in our calculations, with the same size distribution used by Abel et al. (2008). The abundance of carbon atoms in PAHs that we use, $n_C(PAH)/n_H$, is $3\times10^{-6}$. PAHs are thought to be destroyed by hydrogen ionizing radiation and coagulate in molecular environments (see, for instance, Omont 1986, Verstraete et al. 1996). We model this effect by scaling the PAH abundance by the ratio of n(H$^0$)/ $n_H$, using $n_C(PAH)/n_H = 3\times10^{-6}\times [n(H^0)/n_H]$.

## 2.3 Equation of State

The relationship between density, temperature, and pressure is determined by the assumed equation of state. For simplicity, we assume constant pressure, with thermal, radiation, and magnetic pressure being the dominant contributors to the total pressure. Such an equation of state neglects shocks or other time-dependent effects. We also do not include turbulent pressure in the equation of state, although we assume a small turbulent broadening of 1 km s$^{-1}$ for its effects on line optical depths.

At the illuminated face, which marks the beginning of the H$^+$ region, the pressure is set by the thermal and radiation pressure. We set the initial $n_H$ at illuminated face equal to $10^3$ cm$^{-3}$. The actual densities in the ionized regions of ULIRGs are quite uncertain due to the lack of observational data on IR fine-structure lines which are sensitive to density (such as [O III] 52, 88 µm). Since $U$ is the dimensionless ratio of photon to particle density, fixing the density at the illuminated face means that changes in $U$ directly correspond to changes in $\phi_H$, with $\phi_H$ ranging from $3\times10^9$ to $1.2\times10^{13}$ cm$^{-2}$ s$^{-1}$. The density, $U$, and SED determine the gas temperature and radiation pressure at the illuminated face. As $U$ increases, the force of radiation pushing on the gas becomes increasingly more important. For $U = -0.4$, radiation pressure dominates in the ionized gas. We assume a magnetic field ($B_0$) in the ionized gas of 300 µG.. We use the density-magnetic field relationship given in Crutcher (1999):

$$B = B_0 \left( \frac{n_H(depth)}{n_H(face)} \right)^\kappa \qquad (2)$$

with κ = 2/3, corresponding to conservation of magnetic flux (Crutcher 1999). In equation 2, $n_H$(depth) corresponds to the density at a certain depth in the cloud, while $n_H$(face) is the density at the illuminated face. Choosing $B_0$ = 300μG leads to a magnetic field in the PDR of ~500 - 2000μG, a result consistent with magnetic fields in the ISM of ULIRGs as measured through the observed Zeeman splitting of OH megamaser emission (Robishaw, Quataert, & Heiles 2008). In all models, at $A_V$ = 100 mag, magnetic pressure dominates over thermal and radiation pressure. Figure 1 shows how the density at the face of the atomic gas, $n$(PDR), varies with $U$ and choice of SED. Given our equation of state, $n$(PDR) at the beginning of the PDR is also roughly equal to the density at $A_V$ = 100 mag, for a given value of $U$. The value of $n$(PDR) does increase with $U$ due to the increased effects of radiation pressure (see Pellegrini et al. 2007).

# 3 Results

## 3.1 Chemical Structure

Figures 2 and 3 show how the chemical structure depends on $U$ and SED type. Both plots show the H, C, O, and Ne ionization structure for $U$ = $10^{-1.2}$ and $U$ = $10^{-2.8}$. Figure 2 uses an AGN SED, while Figure 3 shows the results of our calculations for a starburst SED. Both are plotted as a function of $A_V$, as is typical in PDR/XDR calculations.

The effects of dust can be seen by comparing the location of the hydrogen ionization front in the low and high $U$ calculations. The size of the $H^+$ region is significantly larger in the high-$U$ model, as expected from photoionization theory. However, even though the flux increased by a factor of 40, dust absorption of hydrogen ionizing photons in the $H^+$ region caused the distance to the $H^+$ ionization front to increase by only a factor of 15. In the absence of dust, the volume of the $H^+$ region increases linearly with $U$. This shows one important consequence of high $U$. The increased dust absorption in the $H^+$ region leads to fewer photons available to ionize hydrogen, leading to a smaller $H^+$ region than would occur in the absence of dust.

The ionization structure in the high and low $U$ cases differs in important ways. As expected, the abundance of higher stages of ionization in the $H^+$ region is greater for high $U$. In the low-$U$, starburst model, the dominant stages of oxygen are $O^+$ and $O^{2+}$, while for high $U$ all the O is in $O^{2+}$. The same is true for Ne, where the low $U$ model predicts all Ne in the form of $Ne^+$ and $Ne^{2+}$ and the high $U$ model predicts all Ne in $Ne^{2+}$. This effect is more extreme for the AGN

SED. For low $U$, almost all O is in $O^{2+}$, but for high $U$ all O is in $O^{3+}$ and higher ionization states. Almost all neon is in $Ne^{2+}$ and $Ne^{3+}$, while for high $U$ $Ne^{4+}$ and $Ne^{5+}$ are the dominant ionization stages. It is important to note that, for a given ionization stage of an element, the ionization fraction does not increase monotonically with increasing $U$. Figure 4 shows the ionization fraction of $Ne^+$ – $Ne^{5+}$ as a function of $U$ at the illuminated face of our calculation. For a given stage of ionization, the ionization fraction of Ne increases with increasing $U$, peaks at a particular value of $U$, then decreases by several orders of magnitude. Gas with $U = 10^{-2.4}$ has $Ne^{3+}$ as the dominant ionization stage, while for $U = 10^{-1}$ $Ne^{5+}$ is the dominant ionization stage and $Ne^{3+}$ has only a trace abundance. Additionally, for a given $U$, the order of abundance does not necessarily go from higher to lower ionization states. For instance, for $U = 10^{-2.5}$, the abundance of ionization stages of neon (in order of highest to lowest abundance) is $Ne^{3+}$, $Ne^{2+}$, $Ne^{4+}$, $Ne^{5+}$, $Ne^+$. Of course, the dominant stage of ionization changes with depth through the cloud as well, with lower ionization states becoming more prominent with increasing depth. This complicated dependence of the level of ionization on $U$ is important in interpreting the emission line spectrum, as we describe in Section 3.2.

Changes in $U$ lead to changes in the dust temperature ($T_{dust}$), since $U$ scales with flux, given our equation of state. Figure 5 shows the predicted $T_{dust}$ of a 0.15μm graphite grain at the illuminated face of our calculation and at $A_V = 100$ mag, for a starburst SED (the results for the AGN model are nearly identical). At the illuminated face, $T_{dust}$ varies from 30 – 160 K, while at $A_V = 100$, $T_{dust}$ ranges from 10 – 60 K. The value of $T_{dust}$ has a significant effect deep in the cloud. For low $U$, the dust temperature is low enough that molecules such as $H_2O$ and CO condense onto dust. For $U < 10^{-2.8}$, most of the gas-phase oxygen has condensed into water ice on grain surfaces. This can be seen in the C and O ionization structure plots in Figures 2 and 3. For low $U$, there is a decrease in the CO and $O^0$ abundance at high $A_V$ which is due to $H_2O$ freeze out. The starburst O structure clearly shows two dips in the $O^0$ structure at high $A_V$. The first dip is caused by CO formation, while the second is a result of $H_2O$ condensation, which is known to occur at a higher temperature than CO. At high $U$ $T_{dust}$ is high and the rate of thermal evaporation of molecules is much higher than the rate of molecular condensation on grain surfaces. Of course, some ULIRGs are observed to have strong ice absorption, which from this discussion would seem to eliminate high $U$ for those galaxies. However, our models were all chosen to stop at a fixed $A_V = 100$ mag. In reality the $A_V$ of ULIRGs will span a wide range, so if ULIRGs are characterized by high $U$ then our models suggest those ULIRGs with ices will likely have a larger column density.

The H→$H_2$ and $C^+$→$C^0$→CO transition is much broader in the AGN than the starburst. This is due to the AGN emitting X-rays which penetrate beyond the $H^+$ region. The atomic/molecular region surrounding the AGN is an XDR, while the region around the starburst is a PDR. In the AGN models, hydrogen is never fully converted to $H_2$ at high $U$, while the fraction of $H^0$ stays above 10%. CO does not exceed 10-30% in either the low or high $U$ models, with most C in the form of $C^0$. Additionally, X-rays keep neon partially ionized (1-10%) well beyond the $H^+$ ionization front. The presence of $Ne^+$ in regions where hydrogen is in the form of $H^0$ is also predicted by Maloney (2003) and Glassgold, Najita, & Igea (2007). For the starburst SED models, all hydrogen eventually ends up in $H_2$, and all C ends up in CO. The lack of X-rays leads to far fewer $H^+$ and $Ne^+$ atoms beyond the $H^+$ region. The AGN results are entirely consistent with the ionization structure shown in the XDR results of Meijerink & Spaans (2005), while the PDR results are consistent with the results of Kaufman et al. (1999). This is consistent with the results shown in Röllig et al. (2007) that PDR calculations made by Cloudy are consistent with other PDR codes.

## 3.2 Predicted Spectrum

A series of predicted emission line and line-to-continuum ratios is presented in Figure 6. Figure 6 shows how various diagnostics pertinent to Herschel, SOFIA, and Spitzer vary as a function of SED and $U$.

Figure 6a shows the predicted [C II] 158 µm /FIR ratio. The [C II] 158 µm /FIR ratio decreases with increasing $U$, a trend suggested in L03, but for the first time demonstrated through a detailed calculation in this work. Physically, the reason for the decrease in [C II] 158 µm /FIR in our calculations is that as $U$ increases, dust absorbs a larger fraction of UV photons, leading to less photons available for the gas (see Section 1). Additionally, density and temperature effects also come into play at higher $U$. This is not surprising, as increasing $U$ increases the UV flux, which increases the gas temperature ($T_{gas}$). Additionally, in the higher $U$ models, $n_H$ exceeds the critical density for the $C^+$ $^2P_{3/2}$ level via $H^0$, $H_2$ collisions ($3\times10^3$ cm$^{-3}$, see Figure 1). The AGN SED predicted slightly higher [C II] 158 µm /FIR than the starburst, which is largely a result of the hotter XDR produced by the AGN. Figure 6a shows that, for $10^{-3.2} < U < 10^{-2}$ (starburst) and $10^{-2.5} < U < 10^{-1}$ (AGN), the [C II] 158 µm /FIR ratio falls in the range observed in the L03 ULIRG sample.

Figure 6b and 6c show the ratio of intensities for [O I] 63µm / [C II] 158µm and [O I] 145µm / [C II] 158µm. For our models, most [C II] 158 µm emerges from PDR gas (>80%), due to the low critical density (50 cm$^{-3}$) of $C^+$ excitation via electrons. For both the AGN and starburst, the [O I]/[C II] 158 µm ratios increase with increasing $U$. The increase in $T_{gas}$ increases the emission of the [O I] lines, since the $J = 1, 0$ levels of the $O^0$ ground term require more energy ($E_u/k_b$ = 228 K for $J = 1$, $E_u/k$ = 327 K for $J = 0$) to populate than the $J = 3/2$ state of $C^+$ ($E_u/k$ = 91 K). The critical density of the $C^+$ levels is ~2 dex smaller than either of the critical densities of the $O^0$ levels. For higher $U$, the density in the PDR exceeds the critical density for the $C^+$ levels (see above), while $n_H$ never exceeds the critical densities for the $O^0$ levels. This is consistent with the result of Malhotra et al.

(2001) that as $G_0$ increases, $n$ increases. Therefore, for high $U$, [C II] 158 μm emission does not increase as rapidly as [O I] 63 μm with increasing $n_H$.

For both [O I]/[C II] 158 μm diagnostics, the predicted ratio is larger for the AGN calculation. The larger line ratios for the AGN are due to the fact that X-rays emitted by the AGN cause the atomic/molecular gas to be hotter and more ionized than the UV dominated starburst SED. The differences between the AGN and starburst calculations are consistent with the trends seen in the XDR/PDR calculations of Meijerink & Spaans (2005), only our calculations include the adjacent $H^+$ region.

In Figure 7, we plot the predicted AGN to starburst ratio of $O^0/C^+$ column densities and the average $H_2$ temperature (temperature weighted by the $H_2$ abundance and averaged over the thickness of the model) as a function of $U$. Figure 7 shows that both the ratio of $N(O^0)/N(C^+)$ and the temperature are higher for the AGN, explaining why the [O I]/[C II] 158 μm ratios are higher for the AGN. The $N(O^0)/N(C^+)$ ratio being higher is primarily due to the lack of CO produced in the AGN model compared to the starburst (see Figure 2 & 3), which increases the $O^0$ column density. The AGN temperature is higher because of X-rays penetrating deep in the cloud and heating the gas through photodissociation and photoionization. For these reasons, the AGN heated XDR yields a higher temperature than the FUV and cosmic ray heated starburst.

Figure 6d shows the ratio of the [C I] emission lines at 370μm and 610μm. The [C I] ratio increases with increasing $U$, with the ratio predicted from the AGN SED again being higher than the starburst. This is a consequence of the much larger $C^0$ region produced by X-rays from the XDR, and the higher $T_{gas}$ produced in the XDR. The values of $E_u$ for the $^3P_1$ (the upper level for [C I] 610μm emission) and $^3P_2$ (the upper level for [C I] 369μm emission) are 24 and 62 K, respectively. Since the $C^0$ region is hotter in the AGN model, the [C I] 369μm line emission will be higher relative to [C I] 610μm emission in the AGN model than the starburst calculations. Figure 6d agrees very well with the medium density XDR and PDR models presented in Meijerink, Spaans, & Israel (2007). Additionally, in the [C I] ratio, the abundance/column density cancels out.

Figure 6e shows the predicted ratio of [O III] 88μm/FIR. The [O III] 88 μm /FIR ratio reaches a maximum and falls off for both the starburst and AGN, with the AGN [O III] 88 μm /FIR ratio peaking at a smaller value of $U$. The shape of the [O III] 88 μm /FIR curve is due mostly to changes in the oxygen ionization balance, an effect already noted for neon in Figure 4. The fraction of O in the form of $O^{2+}$ at the illuminated face peaks around $U = 10^{-3}$, and then starts to decrease again as $O^{3+}$ becomes more prominent. The larger number of $O^{2+}$ ionizing photons produced by the AGN allows $O^{3+}$ to be efficiently produced at lower $U$ than the starburst SED. The starburst SED does eventually produce $O^{3+}$ ions for $U > 10^{-2}$, which agrees with [O IV] 25.88μm observations of starbursts seen by Lutz et al. (1998). For $U > 10^{-2}$, dust absorption of $O^+$ ionizing photons also plays a role in reducing the [O III] 88 µm intensity relative to the FIR.

Figure 6f shows the ratio of [N III] 57µm/[N II] 121µm. Just like [O III]/FIR, this ratio reaches a peak for the AGN SED, and then decreases as $N^{2+}$ becomes less abundant than $N^{3+}$. The same is not true for the starburst calculation. Figure 8 shows the fraction of N in the form of $N^+$, $N^{2+}$, and $N^{3+}$ at the illuminated face. Figure 8 shows that $N^{2+}$ is the dominant ionization state for $U = 10^{-3}$ to $U = 10^{-0.8}$, and only starts to decrease at the highest values of $U$ considered ($U = 10^{-0.4}$). Because of the softer starburst continuum, the [N III]/[N II] ratio increases with $U$, only starting to flatten out at highest $U$. Therefore, for high $U$, the starburst SED predicts a much higher [N III]/[N II] ratio. The ratio depends strongly on $U$ and the choice of SED.

Figure 6g gives the predicted [Ne V] (14µm) / FIR as a function of $U$ for the AGN models only, since the starburst SED typically produces little $Ne^{3+}$ ionizing radiation (Abel & Satyapal 2008). For increasing $U$, [Ne V] 14 µm /FIR increases as a result of the increasing number of $Ne^{4+}$ ions generated by the AGN, peaking at $U = 10^{-1.8}$. For higher $U$, several factors come into play. Higher $U$ results in a larger fraction of $Ne^{3+}$ ionizing photons being absorbed by dust. Additionally, the increasing level of ionization leads to more $Ne^{5+}$ relative to $Ne^{4+}$. Both factors act to decrease [Ne V] 14 µm emission relative to the FIR. However, for high $U$, the density in the $H^+$ region increases as a result of increased radiation pressure. Since the critical density for $Ne^{4+}$ $^3P_1 \rightarrow ^3P_2$ is $> 10^5$ cm$^{-3}$, well above the densities considered in our calculations, the increase in density increases [Ne V] 14 µm emission. There is also an increase in the temperature of the $H^+$ region at high $U$. These factors keep the [Ne V]/FIR ratio relatively constant (~$10^{-3}$) for $U > 10^{-2}$. A similar effect is demonstrated in Gorjian et al. (2007) for the [Ne V]/[Ne III] 15.6µm ratio.

Figure 6h shows the predicted $F(60)/F(100)$ flux ratio. The AGN and starburst calculations produce almost exactly the same variation of $F(60)/F(100)$ with $U$. This is due to the AGN and starburst models producing similar values for $T_{dust}$, as mentioned in Section 3.1. As $U$ increases, $F(60)/F(100)$ increases because we are changing $U$ by changing the flux. Since the dust temperature increases with increasing flux, the dust emits energy at shorter wavelengths and increases $F(60)$ relative to $F(100)$. This effect is further illustrated in Figure 9, which shows the transmitted continuum at $A_V = 100$ mag. For $U = 10^{-1.2}$, the peak of the IR emission occurs at 50µm, which is in the $F(60)$ band. For $U = 10^{-2.8}$, the peak IR emission occurs at ~85µm, which is in the $F(100)$ band. Although not the focus of this work, the SED peak will also vary with $A_V$, with the peak shifting to longer wavelengths, corresponding to cooler dust, for larger $A_V$.

We note that the emission line and line-to-continuum trends in the data are consistent with variations in $U$, a result that is independent of the way in which $U$ is varied. However, the $F(60)/F(100)$ color **is** dependent on the way in which $U$ is varied. In our models, we vary $U$ by varying the flux, with larger $U$ characterized by larger flux. This is a reasonable assumption, as we want to understand the trend in observations of galaxies as the luminosity of the galaxy increases, and luminosity enters into $U$ through the flux. If we had changed $U$ by changing $n_H$ or the inner radius of the model, then we would not see the direct relationship between $U$ and $F(60)/F(100)$.

The effects of dust in the $H^+$ region on the IR spectrum are shown in Figure 10 (where the $H^+$ region is defined as the region where the fraction of all hydrogen atoms that are ionized exceeds 1%). Figure 10 shows the percentage of the total SED intensity which is absorbed by dust in the $H^+$ region. This figure shows the same behavior as seen in Voit (1992), Bottorff et al. (1999), and Abel et al. (2003), which is as $U$ increases, the fraction of the total SED intensity absorbed by dust increases. The AGN has a larger fraction of its total intensity absorbed by dust. This is due to the slightly larger $H^+$ region produced by the AGN, which can be seen by comparing the H ionization structure in Figures 2 and 3. The larger $H^+$ region means a larger dust column density and hence more absorption of UV photons. Figure 10 also shows the percentage of the total SED intensity re-emitted as FIR in the $H^+$ region. Again, the AGN emits more FIR than the starburst, also due to the larger $H^+$ region in the AGN. The difference between Fig 10a and Fig10b gives the fraction emitted from the plasma at wavelengths shorter than 40μm. Comparing Figure 10a to 10b, we find about half of the IR continuum emitted by dust in the $H^+$ region is in the FIR band at low $U$, with the rest of the IR being emitted shortward of 40μm. As $U$ increases, the percentage of FIR also increases. However, for high $U$, the percentage of FIR emitted in the $H^+$ region decreases. This is because, for increasing $U$, $T_{dust}$ increases. As a result, more and more of the total IR continuum is emitted shortward of 40μm, the lower end of the FIR band used in the calculations.

Figure 11 shows how the same emission-line and line-to-continuum ratios vary as a function of $F(60)/F(100)$, instead of $U$ (Figure 6a-g). Since $F(60)/F(100)$ is an observable, Figure 11 is more easily comparable to observations. In Figures 12 & 13 we compare our calculations with observations made with *ISO* and *Spitzer*, while Figure 11 shows calculations for emission-line ratios which will be commonly observed in the future, such as the [C I] ratio and [N III]/[N II].

# 4 Discussion

## 4.1 Observational Database

We have assembled the currently available infrared fine-structure line fluxes and associated continuum fluxes for nearby galaxies in order to compare several key line and line-to-luminosity ratios with the predictions of our generic model and to assess whether the hypothesis suggested in this work is supported by the trends and order of magnitude values of these ratios in the local universe. Since the IRAS band ratio $F(60)/F(100)$ is a monotonic function of $U$ (when $U$ is varied through variations in flux, and neglecting the dependence on stopping criterion), with little dependence on the SED shape (as discussed in section 3.2 and shown in Figure 6h) we use it as an observational measure of the ionization parameter for this assessment. One important observational constraint on infrared line flux ratios and line-to-bolometric luminosity ratios is that such comparisons are only valid where aperture effects are small or negligible. In what follows, we therefore concentrate on far-infrared line fluxes obtained with the *ISO Long Wavelength Spectrometer (LWS)*, whose circular aperture is approximately 70 – 80 arc seconds in diameter across the full 43 – 197µm range, equal to or larger than the emission regions of many of the local galaxies whose far-infrared luminosities were measured by the *IRAS* survey. We also include [Ne V] 14µm in the dataset based on the assumption that the [Ne V] region is compact compared with the *Spitzer IRS* slit (4.7" by 11.3" for the SH setting). For ULIRGs, the *LWS* line fluxes were assembled from Luhman et al. (2003), Fischer et al. (1997), and Gonzalez-Alfonso et al. (2008) and for less luminous galaxies they were taken from the compendium by Brauher et al. (2008) for galaxies listed there as spatially unresolved by the *LWS*. The [Ne V] 14 µm line fluxes were assembled for AGN with redshift $z$ < 0.1 from Schweitzer et al. (2006), Armus et al. (2006), Dudik et al (2007), Farrah et al. (2007), Tommasin et al. (2008), and Dasyra et al. (2008). The IRAS 60 µm and 100 µm fluxes used are those listed in the Revised Bright Galaxy Sample (Sanders et al. 2003), in the IRAS Faint Source Catalogue, or in Brauher et al. (2008). The 40 - 122.5 µm FIR fluxes are computed according to the prescription given in Appendix B of Catalogued Galaxies and Quasars Observed in the IRAS Survey (1985).

## 4.2 Comparison to Observations

Figure 12 displays, for both the starburst and AGN SEDs, the model predictions of the ratios [C II] 158 µm /FIR, [O I] 63 µm /[C II] 158 µm, [O I] 145 µm /[C II] 158 µm, and [O III] 88 µm /FIR. Figure 13 shows the [Ne V] 14 µm /FIR ratio (AGN only). Both Figures 12 and 13 are plotted as a function of $F(60)/F(100)$, and show both the calculated and observed flux ratios. We denote as AGN all galaxies classified exclusively as Seyferts by the NASA/IPAC Extragalactic Database (NED). Since there is considerable variation in the optical classification scheme for LINERs, transition objects, and normal galaxies (e.g. Heckman et al. 1980; Veilleux & Osterbrock 1987; Ho et al. 1997), we denote all galaxies classified as "HII", "LINER", "transition objects", or any combination, as normal/starburst galaxies. Giving LINERs a separate classification, or grouping them with the AGN observations, do not affect our results. In Figure 12, AGN are plotted with the AGN models, normal/starburst galaxies are plotted with the starburst models, and ULIRGs with and without AGN, are plotted in together with both AGN and starburst models.

In comparing our model predictions with observations, we note that the parameters for each individual galaxy will vary, and that the adopted parameters in our model are somewhat arbitrary. Therefore, discrepancies between the models in Section 3 and observations will be expected due to our assumptions regarding our stopping criterion ($A_V$ = 100 mag), choice of density in the $H^+$ region ($10^3$ cm$^{-3}$), equation of state, SED parameters, the neglecting of shocks, geometry, viewing angle, and line optical depth effects. For instance, $A_V$ towards Arp 220 may be as high as $10^4$ mag (Gonzàlez-Alfonso et al. 2004), which will lead to a different emission-line and line-to-continuum spectrum when compared to our 100 mag model. Altering the equation of state will lead to different densities in both the $H^+$ region and PDR, which will also have consequences for the emission ratios presented here. Future observations with Herschel and SOFIA will also help to refine the models/trends. Our fundamental goal is to investigate whether the observational *trends* in the sample can be replicated by variations in $U$ using a standard model. Discrepancy between model and observation likely means a broader investigation of the parameter space or the inclusion of other physical processes (such as shocks or multi-dimensionality) is necessary. Overall, while such investigations will be the focus of future investigations, the models and results presented here are an important first step in modeling the coupled $H^+$ region, PDR, and dust spectrum through a single calculation.

The observed $F(60)/F(100)$ ratios in our sample range from 0.3 to 1.5. Compared with our standard model, this corresponds to log $U$ values ranging from -4 up to -1 (see Figure 6h), with optically classified AGN accounting for both the highest and lowest values. This range in $U$ corresponds well to the range in $U$ explored in Veilleux & Osterbrock (1987) and Groves et al. (2006).

As discussed by Voit (1992), Bottorff et al. (1998), and in Section 3.2, ionized regions containing standard Galactic dust begin to be dust-bounded at log $U \approx$ -2. According to our models, a $F(60)/F(100)$ ratio of 1 corresponds to log $U$ of around -1.6. In fact, the model ratios presented in Figure 12 show changes in slope of varying degrees around $F(60)/F(100) \sim 1$. For the 21 ULIRGs in the Revised Bright Galaxy sample (Sanders et al. 2003), the mean $F(60)/F(100)$ ratio is 0.98 with a standard deviation of 0.19.

### *4.2.1 [C II] 158μm /FIR*

The [C II] 158μm /FIR ratio is predicted to decline by over an order of magnitude over the range of observed $F(60)/F(100)$ ratios and to be lower for starbursts than for AGN for a given $F(60)/F(100)$. Figure 12 shows over an order-of-magnitude declining range in the observed [C II] 158μm /FIR range with $F(60)/F(100)$, most dramatically for the ULIRGs, as predicted by the models. Our AGN model is generally much better at reproducing the observations than the starburst model, although both reproduce the observed decline in [C II] 158μm /FIR range with $F(60)/F(100)$. For ULIRGs with AGN, the average values of the observations lie near the predicted values while our starburst model predicts a lower [C II] 158μm /FIR ratio than observed. For non-ULIRG AGN, the measured ratios have high dispersion with average values in order-of-magnitude agreement with predictions. Interestingly, the [C II] 158 μm /FIR deficit found for ULIRGs by Luhman et al. (1998) occurs around $F(60)/F(100) \approx 1$ (see Figure 5 of Luhman et al. 2003).

### *4.2.2 [O I] 63μm /[C II] 158μm*

Both the AGN and starburst models predict significant increases in the [O I] 63μm /[C II] 158μm ratios with $F(60)/F(100)$, although higher ratios for given colors and a steeper increase is predicted by the AGN model. Indeed, observations of this ratio in galaxies with and without optically detected AGN show increases with $F(60)/F(100)$ color (Figure 12) although less of a difference between these classes is seen in the observations than in the predictions, and the AGN model predictions are a factor 2-3 higher than the majority of the optically detected AGN and ULIRGs while the starburst models are generally lower, up to a factor of 20, than the observed ratios for most of the starbursts and ULIRGs. Since in the nearest ULIRG Arp 220, the [O I] 63μm line is seen in absorption (Luhman et al. 2003; Gonzalez-Alfonso et al. 2004), this discrepancy is even more difficult to reconcile. Abundance effects or shock contributions to the [O I] 63μm line emission could be responsible for these discrepancies. *Herschel* observations with the spectral resolution of the Photoconductor and Array Camera and Spectrometer (PACS) may be able to assess the ubiquity of absorption and its role in the [O I] and [C II] line strengths in ULIRGs.

### *4.2.3 [O I] 145 μm/[C II] 158 μm*

Both the AGN and starburst models predict an order of magnitude increase in the [O I] 145 μm/[C II] 158 μm ratio with $F(60)/F(100)$, with higher ratios predicted for AGN. For $F(60)/F(100) = 1$, the AGN model predicts a ratio of 1, while the starburst model predicts 0.05 for this line ratio. The predicted ratio for the AGN model is higher than observed, while the predicted ratio for the starburst models for $F(60)/F(100) = 1$ is in general agreement with observations. The higher predicted ratios for the AGN are a consequence of the physics of XDRs, hotter atomic gas combined with less formation of CO leading to more atomic oxygen. However, given the paucity of observations of this ratio in galaxies, the comparison of observational trends with models will need to await future observations. We emphasize the need to observe the [O I] 145 μm line towards AGN so we can refine XDR models and apply these models ULIRGs, where both AGN and starbursts reside. Herschel observations at moderate resolution with the Photoconductor and Array Camera Spectrometer (PACS) and high resolution with the Heterodyne Instrument for the Far-Infrared (HIFI) will enable a significantly more complete comparison of average values and trends in this line ratio with model predictions.

### 4.2.4 [O III] 88 μm/FIR

Both the AGN and starburst models predict steep rises in the [O III] 88μm /FIR ratios with $F(60)/F(100)$ followed by a reversal and slower decline with color at $F(60)/F(100)$ ~ 0.6-0.7. The AGN model predicts somewhat higher line flux ratios at cooler FIR color and then a steeper decline for warmer FIR color. Observations of AGN plotted in Figure 12 are in general agreement with the model predictions, although there is a paucity of observations, both for ULIRGs in general and for starburst/AGN galaxies with cool colors. Although many of the line flux ratios for the starburst galaxies are in reasonable agreement with the models, some of the starburst galaxies with cool colors show fluxes an order of magnitude higher than the model predictions. With its expected sensitivity, *Herschel PACS* will help assess the viability of the high *U* models in explaining the low line intensities in ULIRGs.

### 4.2.5 [Ne V] 14 μm/FIR:

Figure 13 shows the predicted [Ne V] 14 μm/FIR ratio, only for the AGN calculations. The AGN model predicts the general trend of a steep rise and subsequent leveling off with FIR color at $F(60)/F(100)$ ~ 1, which is indeed observed in the optically-detected AGN data. While the non-ULIRG optically detected AGN tend to lie above the curve and the ULIRGs, which appear as intermediate FIR color objects when compared with the whole AGN sample, lay below it. As a whole the sample line ratios are reasonably centered on the model curve. The lowest $F(60)/F(100)$ galaxies do have observed [Ne V] 14 μm/FIR which appear to be 2-3 dex higher than our model calculations, but most of these observations are upper limits, meaning they could still be consistent with our model. Higher extinction in ULIRGs may be responsible for the lower ULIRG [Ne V] line flux ratios compared with the model and non-ULIRG AGN with comparable FIR color.

## 5 Summary and Conclusions

In this paper, we have predicted the observational consequences of the presence of high incident ionization parameters (*U*), that is, high ratios of incident photon to particle densities, in gas rich galaxies in order to 1) explore the possibility that such "dust-bounded" conditions explain trends seen in data available in the current literature, in particular for ULIRGs and other galactic nuclei with warm IRAS $F(60)/F(100)$ colors and 2) in order to prepare for the interpretation of infrared spectroscopic and imaging data that will become available from large *Spitzer Space Telescope* and *Herschel Space Observatory* surveys. High *U* effects include increasing the number of UV photons absorbed by dust relative to the gas, along with increasing the level of ionization of species such as Ne and O in the H$^+$ region. In our model, since we varied *U* by varying flux, high *U* also corresponds to increasing $F(60)/F(100)$. Rather than model each galaxy individually, we developed two simple constant pressure, spherically symmetric models tuned to the average properties of local galaxies illuminated by 1) an AGN and 2) a starburst SED extending through a single value of the gas column density, varying only a single parameter, the ionizing photon flux $U$. Because the model parameters are somewhat arbitrary and not tuned to each galaxy, some discrepancy between the models and the data is expected; however we demonstrate that our model *trends* are in qualitative agreement with the observational trends, suggesting that an increasing ionization parameter can replicate the observed far-infrared line flux and continuum flux density trends that might be expected to accompany increasing galaxy luminosity. In order to avoid aperture effects as much as possible, we concentrated on far-infrared line fluxes obtained with the *ISO Long Wavelength Spectrometer (LWS)*, whose aperture is equal to or larger than the emission regions of many of the local galaxies whose far-infrared luminosities were measured by the *IRAS* survey, but also included [Ne V] 14 μm in the dataset based on the assumption that the [Ne V] region is compact compared with the *Spitzer IRS* slit.

Based on our generic models of an $H^+$ region adjacent to atomic/molecular gas, and our comparison of these models with the available data, we conclude:

- Dust-bounded effects due to significant absorption of ionizing photons begin at IRAS $F(60)/F(100)$ ratios of about 0.8, corresponding to $U \approx 10^{-2}$. At these values, our models predict about 30% and 80% of the total illuminating flux is absorbed by dust in the ionized region for starburst and AGN SEDs, respectively and about 10% and 30% of the total luminosity is re-emitted from within the ionized region in the far-infrared (λ between 42.5 – 122.5 μm).

- High ionization parameter effects can explain the nearly order of magnitude drop in the [C II] 158 μm /FIR ratio seen in the ultraluminous class of galaxies and other warm galactic nuclei with high IRAS $F(60)/F(100)$ ratio, known as the "[C II] deficit" both for starburst and AGN SED models.

- Such effects also produce increases in the [O I](63)/[C II] 158 μm ratio consistent with the magnitude of the trends found in the literature, although the chosen starburst and AGN SED models underestimate and overestimate the absolute magnitude of the ratios for a given IRAS $F(60)/F(100)$ color, respectively. For the AGN case, we suggest that this could be partially due to line optical depth effects in [O I] 63μm and (to a lesser extent) the [C II] 158μm. Such effects have previously been suggested to explain of ~65% of ISO observations (Liseau, Justtanont, & Tielens 2006), and are also consistent with the FIR spectra of Arp 220 and the compact H II region NGC 6334A (Abel et al. 2007). The AGN models predict strong [O I]145μm line fluxes of roughly comparable strength to the [C II] line, higher than observed for any AGN, although more observations are needed of the [O I] 145 µm in AGN. For starburst galaxies, our models seem to reproduce the observations.

- The models produce a reasonable fit to the trends observed in the [Ne V] 14 µm/FIR ratio for optically detected AGN with $F(60)/F(100) > 0.5$, although ULIRGs are found to have lower observed ratios compared with non-ULIRGs with comparable $F(60)/F(100)$ ratios.
- Our models predict a steep increase and then gradual decline in the [O III] 88 µm /FIR ratio occurring around IRAS $F(60)/F(100) \approx 0.75$ and 0.95 for AGN and starburst models, respectively. Qualitatively the declines at high $U$ are consistent with the lower [O III] 88 µm /FIR ratio seen in Arp 220.

Because most galaxies are composed of many components, some embedded within obscuring dust and some unobscured, and in particular because the unobscured regions are likely to dominate the line emission spectrum, the finding that most of the trends can be reproduced by such simplified models with Av = 100 mag, suggests that high $U$ effects may in fact play an important role in the observed trends. The models presented here are intended to help begin to understand the spectra of infrared-bright galaxies expected to become available from the *Spitzer Space Telescope* archive, with the launch of the *Herschel Space Observatory*, and the beginning of science operations for *SOFIA*.

Acknowledgements: We would like to thank the anonymous referee for their suggestions, which improved this manuscript. NPA would like to thank Gary Ferland for useful discussions related to this work. This material is based upon work supported by the National Science Foundation under Grant No. 0094050, 0607497 to The University of Cincinnati, and the Spitzer Science Center under award 1356415 to the University of Cincinnati. NPA also acknowledges the University of Cincinnati, the University of Kentucky, and Miami University for a generous allotment of time on their respective supercomputing clusters. JF gratefully acknowledges support for Herschel Space Observatory Key Program Science from the NASA Herschel Science Center. Basic research in infrared astronomy at the Naval Research Laboratory is funded by the Office of Naval Research. SS gratefully acknowledges support from NASA grant NAG5-11432.

Luhman, M. L. et al. 1998, ApJ, 504, 11L

Luhman, M. L. et al. 2003, ApJ, 594, 758

Malhotra, S. et al. 1997, ApJ, 491, 27

Malhotra, S., et al. 2001, ApJ, 561, 766

Maloney, P. R., Hollenbach, D. J., & Tielens, A. G. G. M. 1996, ApJ, 466, 561

Martin, P. G. & Rouleau, F. 1991, eua..coll..341M

Meijerink, R., & Spaans, M. 2005, A&A, 436, 397

Meijerink, R., Spaans, M., & Israel, F. P. 2007, A&A, 461, 793

Meyer, D. M., Jura, M., & Cardelli, J. A. 1998, ApJ, 493, 222

Omont, A. 1986, 164, 159

Robishaw, T., Quataert, E., & Heiles, C. 2008, ApJ, 680, 981

Sanders, D. B., Mazzarella, J. M., Kim, D. –C., Surace, J. A., Soifer, B. T.

Savage, B. D. & Sembach, K. R. 1996, ARA&A, 34, 279

Shaw, G., Ferland, G. J., Abel, N. P., Stancil, P. C., & van Hoof, P. A. M. 2005, ApJ, 624, 794

Stacey, G. J. et al. 1999, ESASP, 427, 973

Storey, P. J. & Hummer, D. G. 1995, MNRAS, 272, 41

Tielens, A. G. G. M. & Hollenbach, D. 1985, ApJ, 291, 722

van Hoof, P. A. M., Weingartner, J. C., Martin, P. G., Volk, K., & Ferland, G. J., 2004, MNRAS, 350, 1330

Veilleux, S. & Osterbrock, D. E. 1987, ApJS, 63, 295

Verstraete, L., Puget, J. L., Falgarone, E., Drapatz, S., Wright, C. M., Timmerman, R. 1996, A&A, 315, 337

Voit, G. M. 1992, ApJ, 399, 495

Weingartner, J. C. & Draine, B. T. 2001b, ApJS, 134, 263
20

# 7 Figures

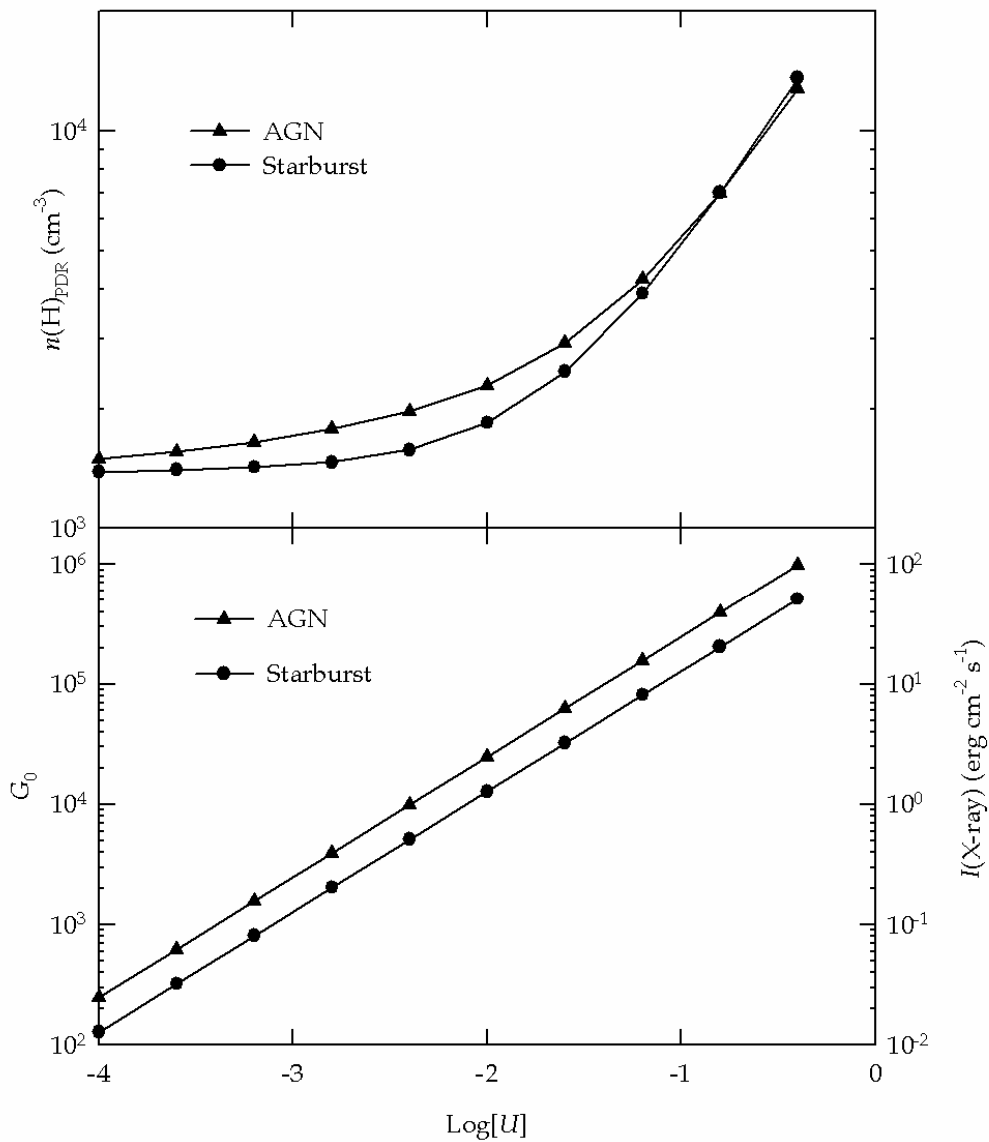

Figure 1 –Density in the atomic/molecular gas as a function of *U* and SED. The density increases with *U* due to the increasing effects of radiation pressure. For $A_V$ = 100 mag, the density increases until magnetic pressure dominates over thermal and radiation pressure. Also shown is the value of $G_0$ (starburst SED only, as $G_0$ for the AGN model is nearly identical) and X-ray intensity (AGN SED only) at the illuminated face, also as a function of *U*.

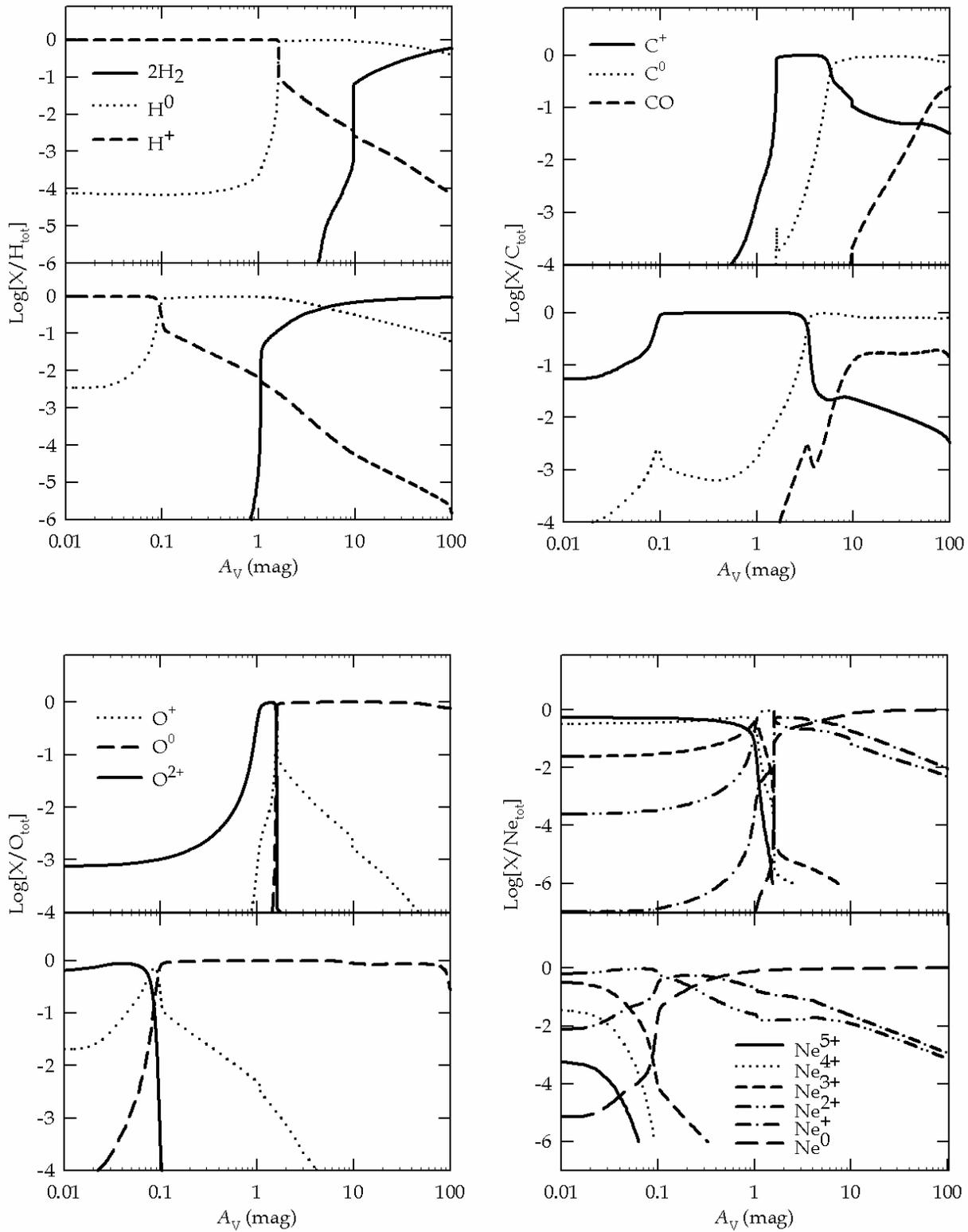

Figure 2 – Chemical structure of H, C, O, and Ne for an AGN SED with an ionization parameter $U = 10^{-1.2}$ (high-$U$; upper panel) and $U = 10^{-2.8}$ (low-$U$; lower panel). The harder radiation field produced ions such as $Ne^{4+}$, while the atomic/molecular gas is characteristic of an XDR.

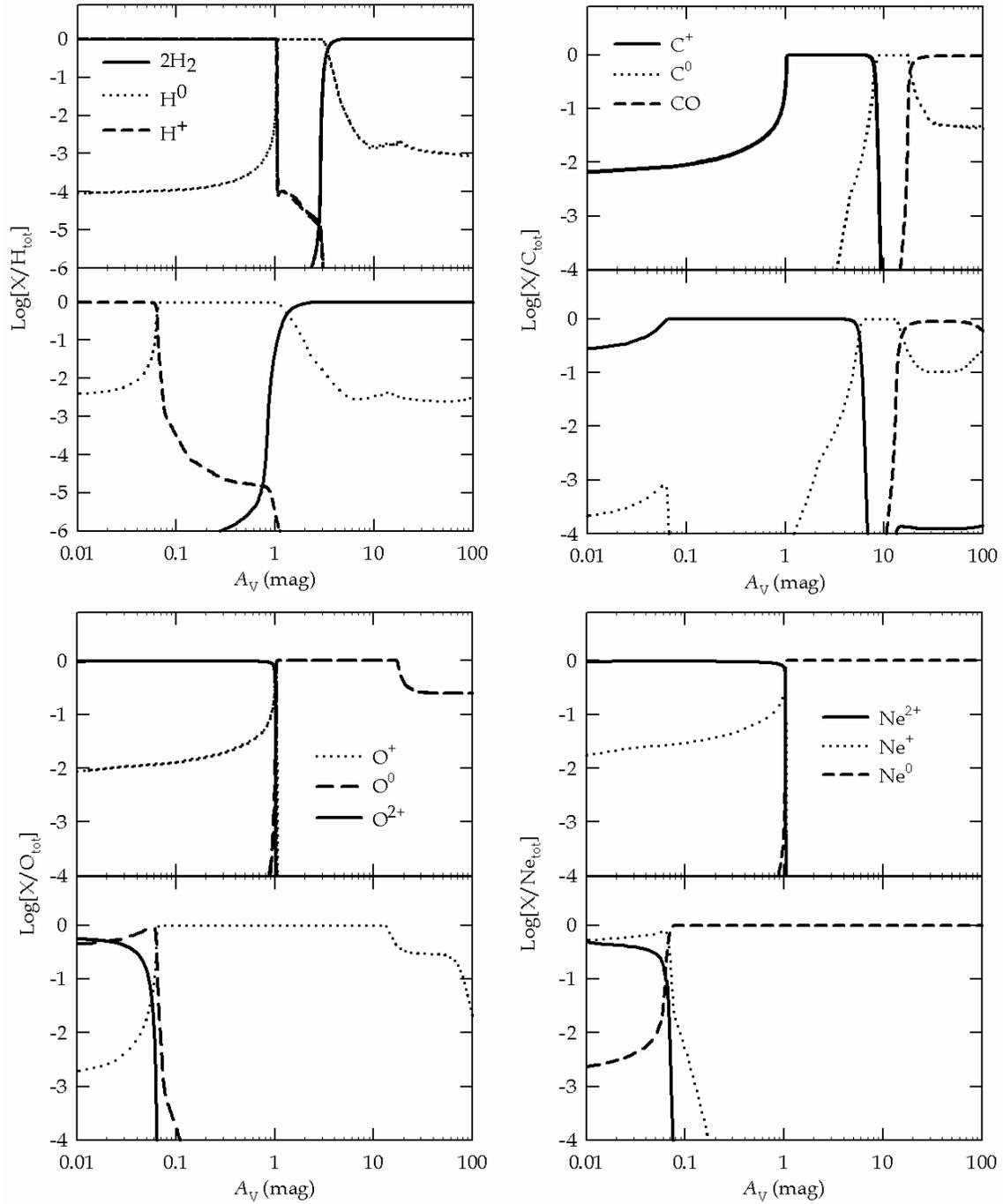

Figure 3 – Chemical structure of H, C, O, and Ne for a starburst SED with an ionization parameter $U = 10^{-1.2}$ (high-$U$; upper panel) and $U = 10^{-2.8}$ (low-$U$; lower panel). Since a starburst produces few photons with energies >54.4eV, ions such as $Ne^{4+}$ are not seen. The surrounding atomic/molecular gas is characteristic of a PDR.

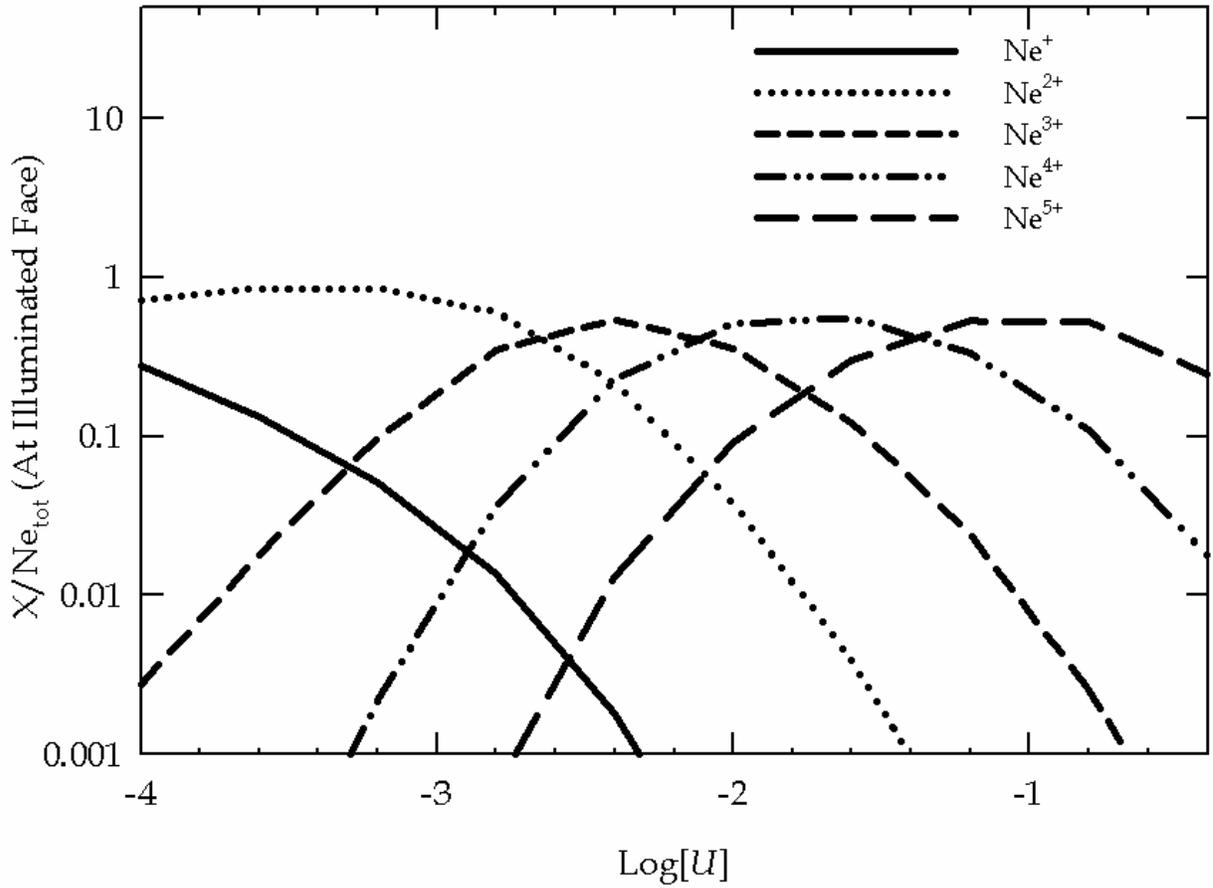

Figure 4 – Ionization fraction of neon at the illuminated face of our calculations, for an AGN SED. As $U$ increases, the dominant ionization stage shifts to higher ionization states. For a given ionization stage, the ionization fraction increases up to a critical value of $U$, then decreases as higher ionization states become more abundant. Ionization effects play a major role in the predicted emission-line spectrum shown in Figures 6 and 11.

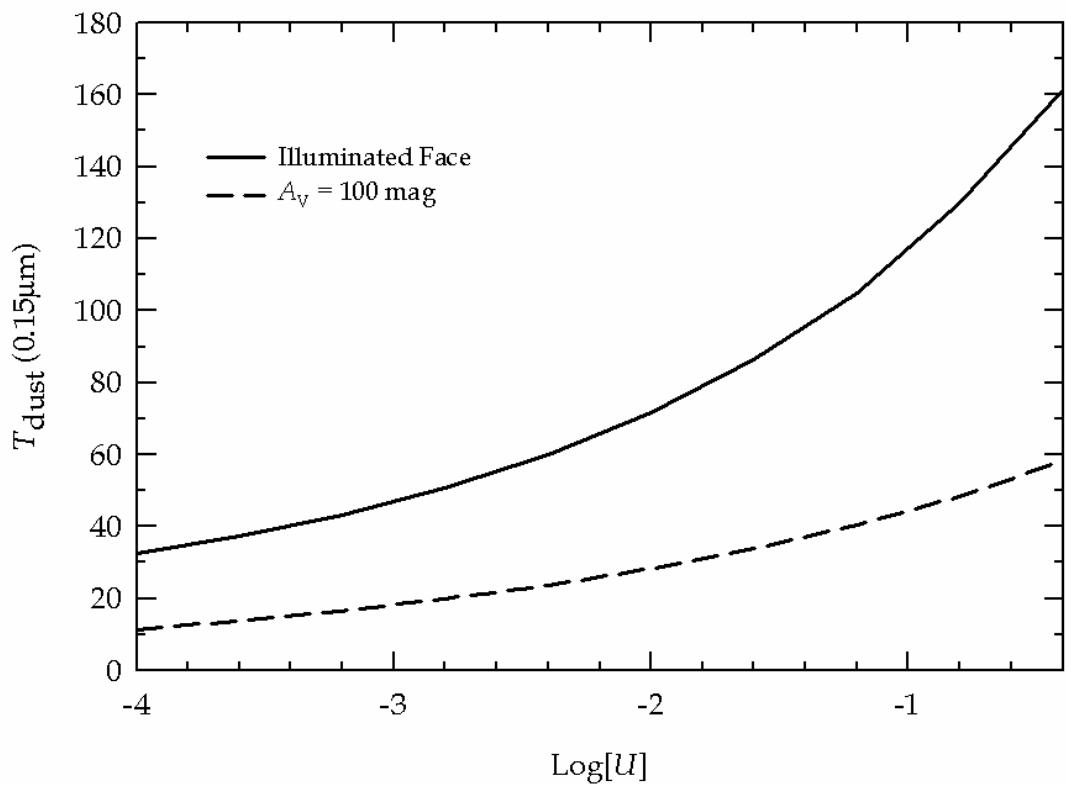

Figure 5 – Dust temperature ($T_{dust}$) of a 0.15μm graphite grain at the illuminated face and at $A_V$ = 100 mag, as a function of $U$. The AGN and starburst models produce nearly identical results, so we only show the AGN results here. As $U$ increases, $T_{dust}$ increases due to the increasing UV flux. The hotter dust temperatures at $A_V$ = 100 mag play a role in the predicted F(60)/F(100) μm flux (see Figure 6) and in the ability of molecules to condense onto grain surfaces.

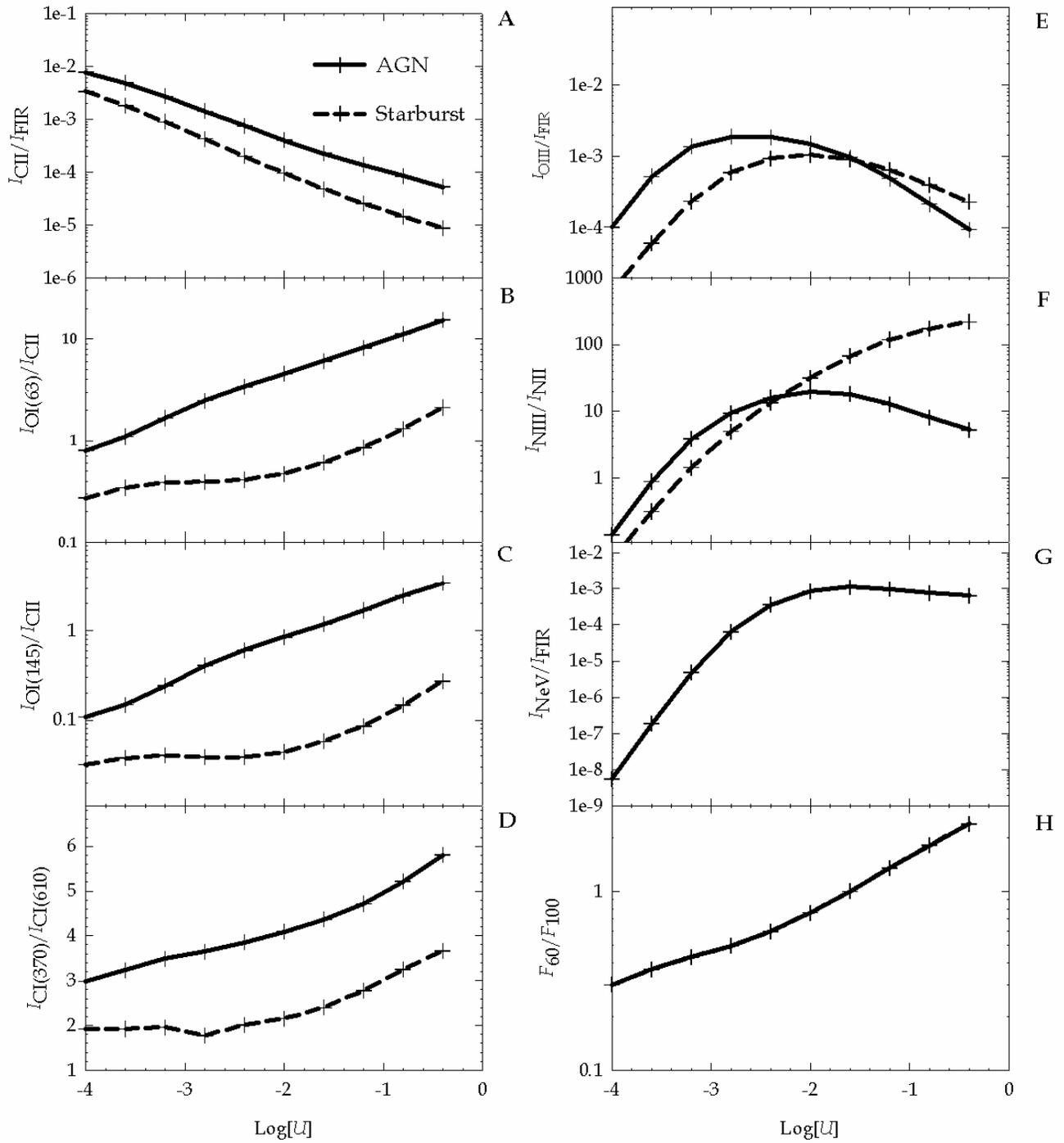

Figure 6 – Predicted emission line and line-to-continuum ratios as a function of $U$ and SED.

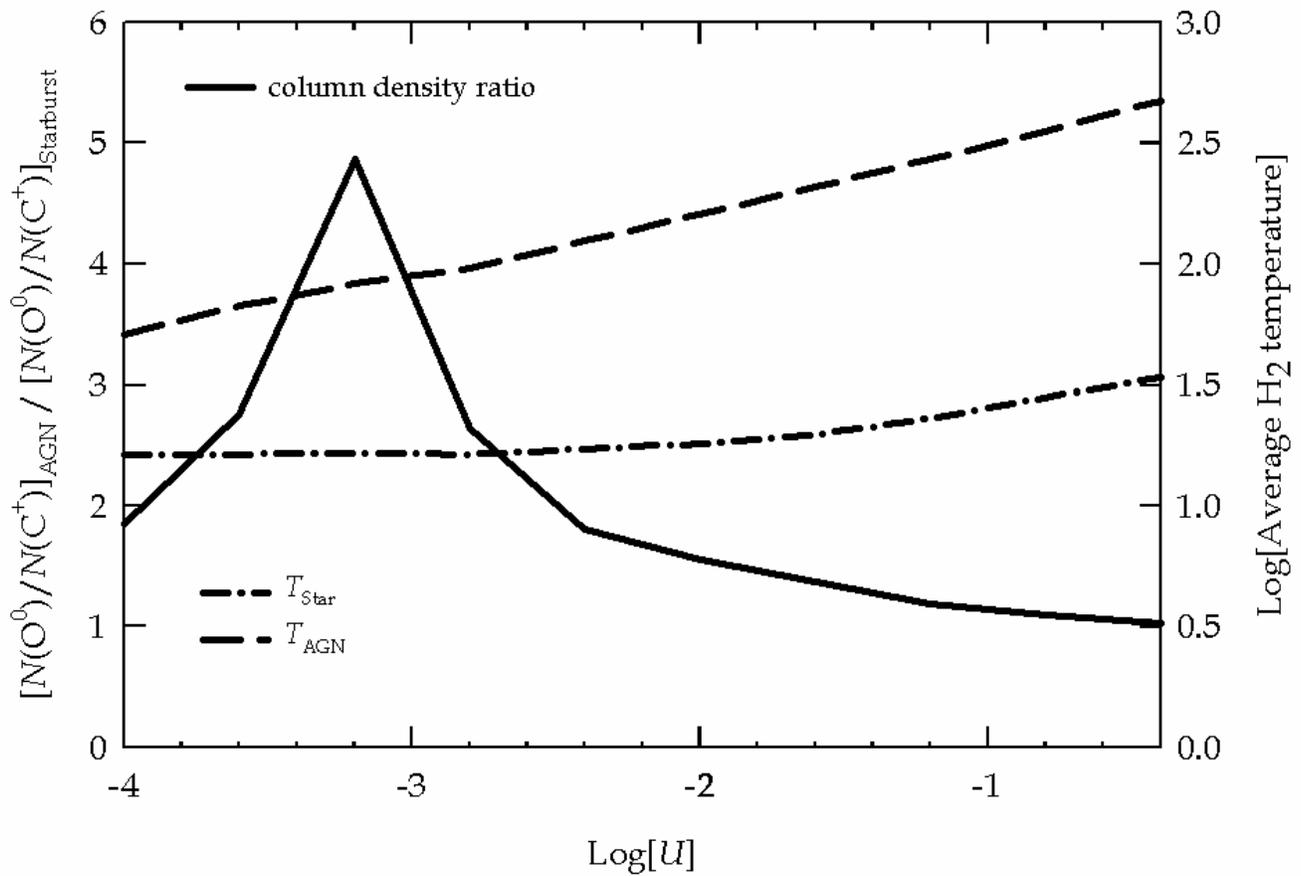

Figure 7 – The ratio of $N(O^0)/N(C^+)$ column densities predicted by the AGN divided by the starburst $N(O^0)/N(C^+)$ column density ratio as a function of $U$. Also plotted is the average $H_2$ temperature for each SED as a function of $U$. This Figure shows why the AGN [O I]/[C II] ratio is higher than the starburst model. The $O^0/C^+$ ratio peaks around $U = 10^{-3.2}$ due to molecules freezing out in the starburst SED, but not for the AGN. This effect reduces the $O^0$ column density in the starburst, as most gas-phase oxygen goes into $H_2O$ on grain surfaces. The most important factor in making the AGN [O I]/[C II] ratio higher is the temperature dependence. Since $T_{AGN} > T_{star}$, the [O I]/[C II] ratio will be higher in the AGN. The higher temperature is the result of X-ray vs. FUV heated gas.

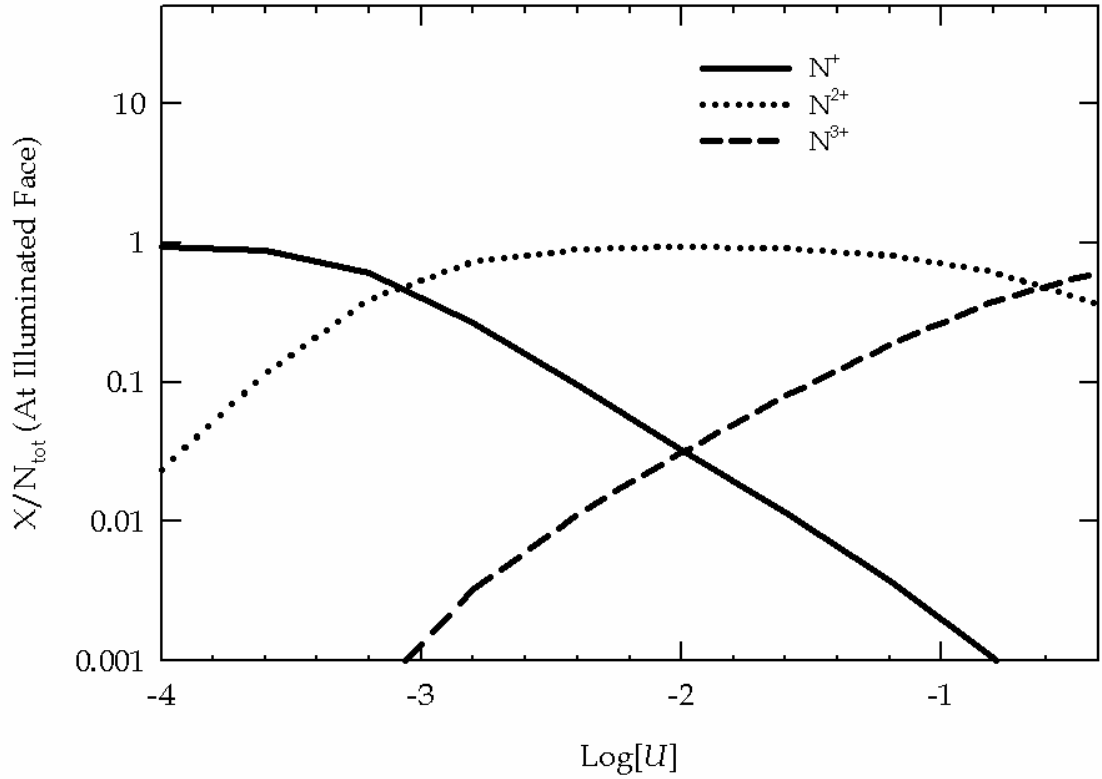

Figure 8 – The ionization fraction of nitrogen at the illuminated face of our starburst calculations, as a function of $U$. Since the starburst SED emits few photons with energies > 54.4 eV, the dominant ionization stages varies less with $U$ than for the AGN SED (See Figure 4). Over a large range in $U$, $N^{2+}$ is the dominant ionization stage. This means that the [N III]/[N II] ratio continues to increase as $U$ increases.

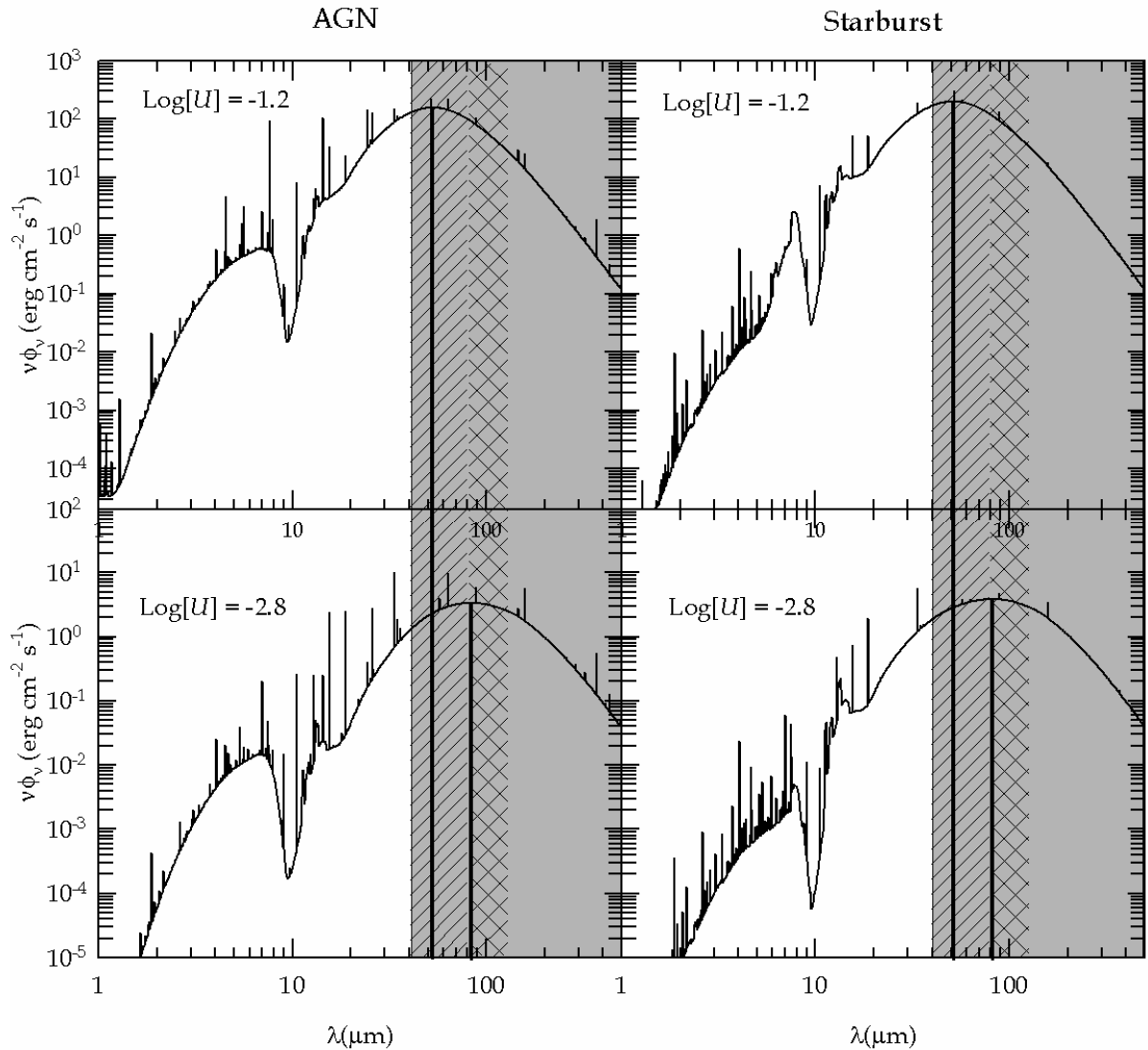

Figure 9 – Transmitted continuum at $A_V$ = 100 mag for a starburst and AGN SED with $U = 10^{-1.2}$ and $U = 10^{-2.8}$. The gray shaded area represents the FIR band, the diagonally shaded region is the 60 μm band, and the cross shaded region is the 100 μm band. The vertical lines point to the peak of the IR emission. As $U$ increases, the dust is heated to a higher temperature, leading to the peak IR emission shifting to shorter wavelengths. For $U = 10^{-1.2}$, the IR peak is in the $F_{60}$ band, while for $U = 10^{-2.8}$ the peak IR shifts to the $F_{100}$ band. The dips seen at 10 and 18μm are due to amorphous silicate dust.

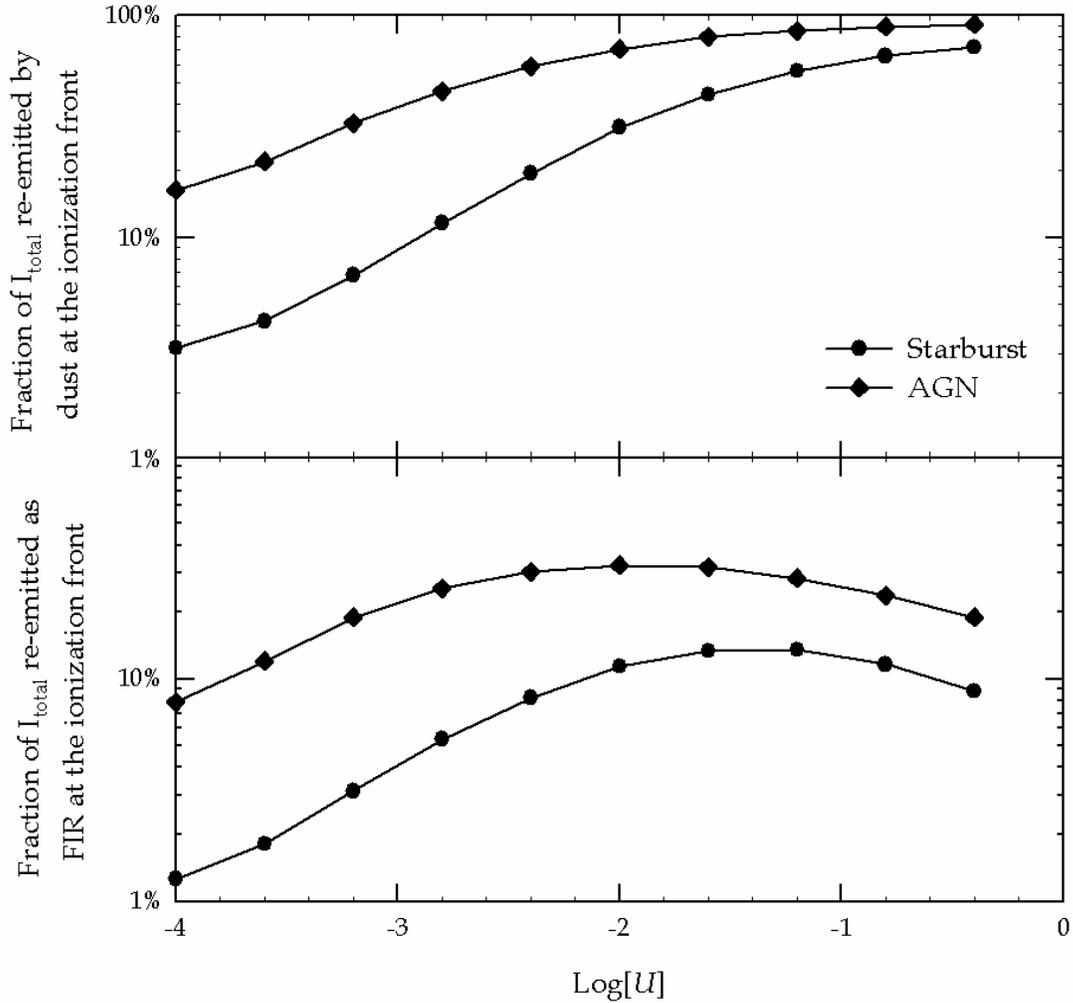

Figure 10 – Fraction of total SED intensity absorbed by dust in the H+ region, along with the fraction of total intensity re-emitted in the 40 – 500 μm range (FIR). As $U$ increases, the dust column density in the H+ region increases, which consequently absorbs more UV radiation. The FIR emitted in the H+ region also increases with $U$, but starts to decrease once the H+ region dust becomes hot enough that most of the emitted IR is shortward of 40μm. The higher fractions from the AGN are due to differences in the size of the H+ region predicted by the AGN and starburst calculations (see Figures 2 and 3).

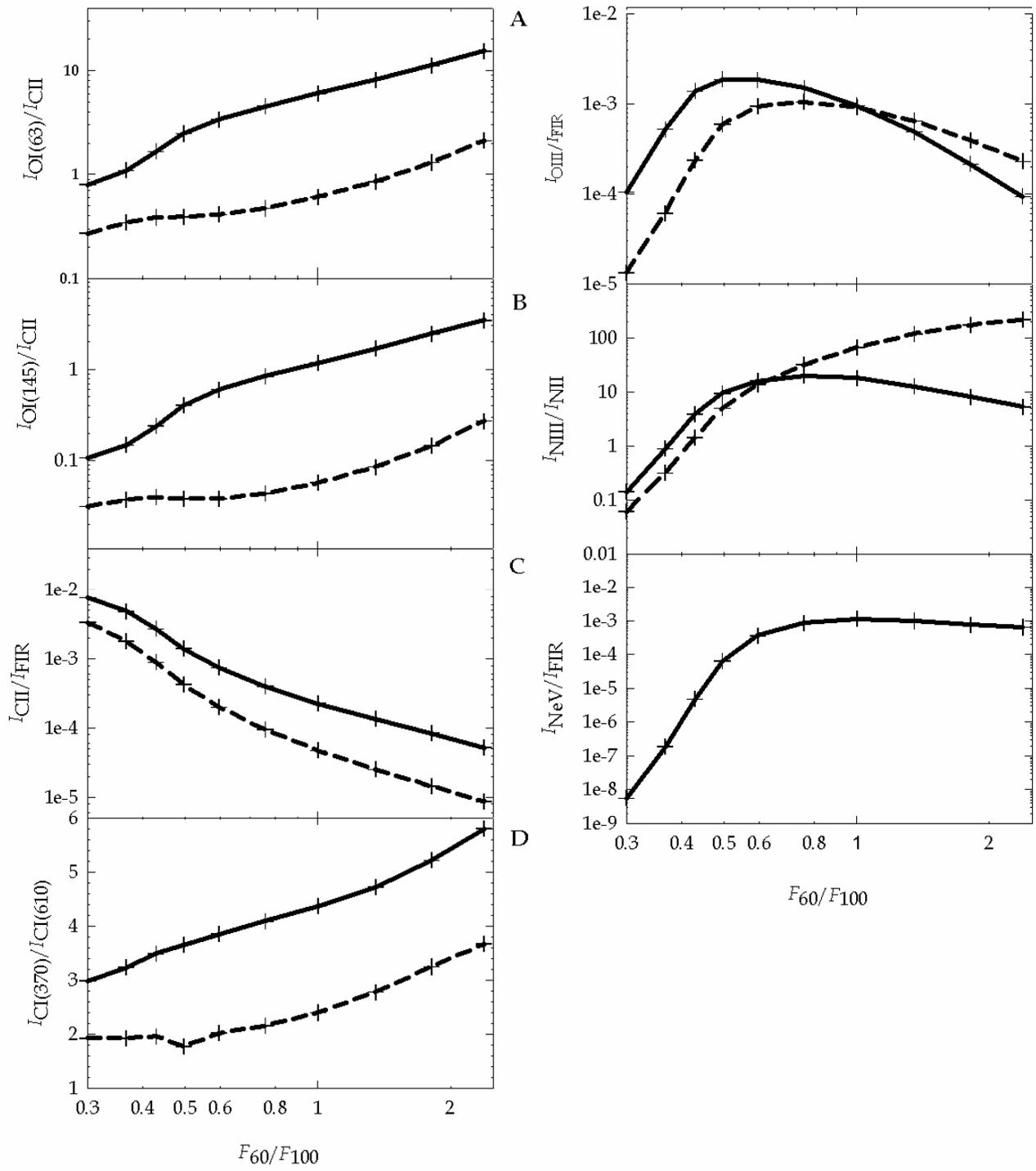

Figure 11 – Predicted emission line and line-to-continuum ratios as a function of $F(60)/F(100)$ and SED.

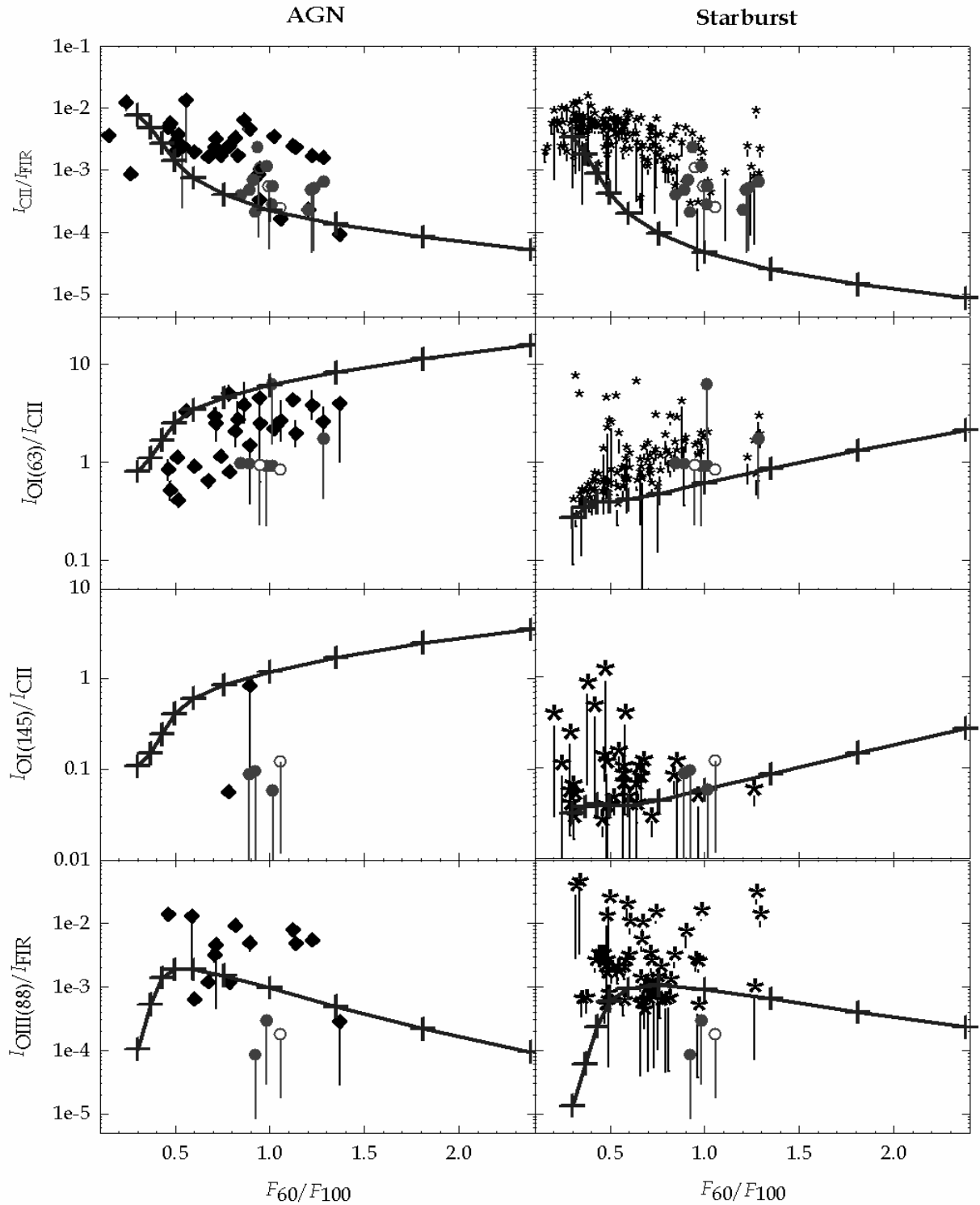

Figure 12 – Comparison of our AGN and starburst calculations with ISO and Spitzer observations. The [C II]/FIR, [O I] (63μm)/[C II], [O I] (145μm)/[C II], and [O III] (88 μm)/FIR is plotted versus $F(60)/F(100)$ (which given our assumptions, is related to $U$) for both the AGN and starburst SED. ULIRGs are plotted with filled (with AGN) and open (no AGN signatures) circles, respectively; AGN in lower luminosity galaxies are plotted with filled diamonds and non-ULIRG normal and starburst galaxies are plotted with asterisks. Vertical bars represent observations with upper or lower limits.

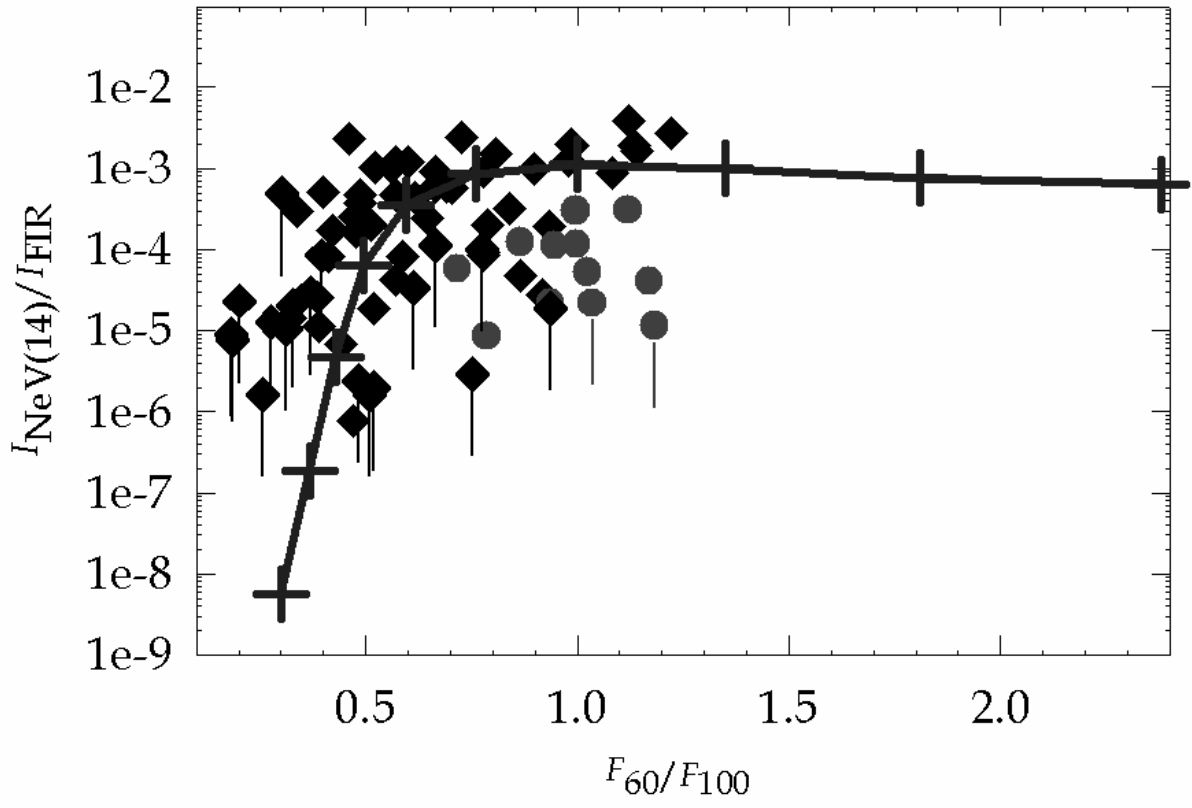

Figure 13 – Comparison of our AGN calculations of [Ne V] 14μm/FIR as a function of $F(60)/F(100)$ with available observational data. The symbols are the same as in Figure 12.

Table 1 – Model Parameters

| Variable | Assumed Values |
|---|---|
| SED | - AGN ($T = 10^6$ K, $\alpha_{ox} = 10^{-1.4}$, $\alpha_{uv} = 10^{-0.5}$, $\alpha_x = 10^{-1.0}$)<br>- Starburst (4 Myr Continuous, power law index of 2.35 star-formation rate = 1 $M_\odot$ yr$^{-1}$) |
| Total hydrogen density at illuminated face of H$^+$ region ($n_H$) | $10^3$ cm$^{-3}$ |
| Initial Magnetic Field ($B_0$) | 300 µG |
| Magnetic Field – Density Relationship | $B = B_0 \left( \dfrac{n_H(depth)}{n_H(face)} \right)^\kappa$ µG; κ = 2/3 |
| Equation of State | isobaric (gas, magnetic, radiation) |
| Gas Phase Abundance (X/H) | ISM Abundances as defined in Cloudy (Section 2.2):<br>- Cowie & Songaila (1986)<br>- Savage & Sembach (1996)<br>- Meyers et al. (1998) |
| Dust Properties | $R_V$ = 4.3; $A_V/N(H_{tot})$ = 5×10$^{-22}$ mag cm$^2$; size distribution from Weingartner & Draine (2001b), using astronomical silicate and graphite (Martin & Rouleau 1991). |
| PAH Properties | Size distribution from Abel et al. (2008), abundance: $n_C(PAH)/n_H$ =3×10$^{-6}$× [$n(H^0)/n_H$] |
| Ionization Parameter ($U$) | From 10$^{-4}$ to 10$^{-0.4}$, in increments of 0.4 dex |
| Cosmic Ray Ionization Rate ($\zeta_{cr}$) | 5×10$^{-17}$ s$^{-1}$ |
| Stopping Condition | $A_V$ = 100 mag ($N(H_{tot})$ = 2×10$^{23}$ cm$^{-2}$) |